# The relationship between burnout operators with the functions of family Tehran Banking Melli Iran Bank in 2015

**Authors:**

- Mohammad Heydari, "Corresponding Author", Master of Business Administration in International Business student orientation, Payame Noor University, International Center of Assaluyeh, Iran, Postcode: 3168144444, E - mail: MohammadHeydari1992@yahoo.com

- Matineh Moghaddam, Master of Business Administration in International Business student orientation, Payame Noor University, Unit of Tehran West, Iran. E - mail: matineh_Moghaddam@yahoo.com

- Habibollah Danai, PhD in Business Administration and Professor, Payame Noor University, Tehran, Iran. E - mail: h.danaei@live.com

# The relationship between burnout operators with the functions of family Tehran Banking Melli Iran Bank in 2015

Mohammad Heydari[1], Matineh Moghaddam [2], Habibollah Danai[3]

**Abstract:**

In this study, the relationship between burnout and family functions of the Melli Iran Bank staff will be studied. A number of employees within the organization using appropriate scientific methods as the samples were selected by detailed questionnaire, and the appropriate data is collected burnout and family functions. The method used descriptive statistical population used for this study consisted of 314 bank loan officers in branches of Melli Iran Bank of Tehran province and all the officials at the bank for more than five years of service at Melli Iran Bank branches in Tehran, They are married and men constitute the study population. The Maslach Burnout Inventory in the end internal to 0/90 alpha emotional exhaustion, depersonalization and low personal accomplishment Cronbach alpha of 0/79 and inventory by 0/71 within the last family to solve the problem 0/70, emotional response 0/51, touch 0/70, 0/69 affective involvement, roles, 0/59, 0/68 behavior is controlled. The results indicate that the hypothesis that included the relationship between burnout and 6, the family functioning, problem solving, communication, roles, affective responsiveness, affective fusion, there was a significant relationship between behavior and the correlation was negative. The burnout is high; the functions within the family will be in trouble.

**Keywords:** burnout, family, emotional response, emotional fusion

[1] Master of Business Administration in International Business student orientation, Payame Noor University, International Center of Assaluyeh, Iran. Postcode: 3168144444, Telephone number (09195286393, 021_65350056), E - mail: MohammadHeydari1992@yahoo.com "Corresponding Author"
[2] Master of Business Administration in International Business student orientation, Payame Noor University, Unit of Tehran West, Iran. E - mail: matineh_Moghaddam@yahoo.com
[3] PhD in Business Administration and Professor, Payame Noor University, Tehran, Iran. E - mail: h.danaei@live.com

## 1. Introduction:

Stress in the workplace is common and many people are facing; therefore, a group of experts in organizational behavior, stress the long-term disease that can lead to century have described it burn out after one of the most work stresses is unavoidable consequences to their physical exhaustion, changes in behavior and job performance brought, in fact, known as burnout and symptoms below:

- Indicators of emotional symptoms such as lack of interest in getting to their jobs, depression, feelings of helplessness and impotence, loss of empathy towards others
- Indicators of attitude of suspicion towards others, to manage their workplace grumbling from other organizations
- Behavioral indices of irritability and aggression, increased drug abuse and rate of complications and conflicts with subordinates, bosses, colleagues, spouses and children
- Psychosomatic factors such as headaches, digestive disorders and ........
- Organizational measures, including theft, absenteeism, reduced employee morale and spiritual dimensions. (Saatchi, 1997)

Person who experience job burnout is a negative impact on their colleagues by creating personal conflicts and interfere in the conduct of their duties and so can be contagious for informal interactions continue in the job also among the works that burnout is a negative spread is spread and its impact on people's lives. (Burke & Greenglass 2001)

Maslach, C., Jackson, S.E and Leiter, M.P. (1996) observed that the level of emotional exhaustion related to the employment of people with spouses reported on the nervous, upset and tired and depressed home they have a valid connection. The emotional exhaustion associated with poor-quality jobs with family life by the couple reported a strong relationship.

About 50 years ago a sociologist Willard Waller called a couple of positions and noted its impact on marriage and family relations (Aygman 1986) He pointed out that the family economic situation deeply affected the process of family conflict. Harsh working conditions and stress due to be transferred to home and husband and wife against each other on stimulated. The high rate of employee separation of family businesses that have many challenges and conflicts, have been observed. A person who sees himself failed in his job, at home and aggression may amend it seems, and it might be noted that such families will not succeed. (Rasuli, 2002)

Family and work are two important aspects of life and the experiences of each other affects on family and work, researchers at describing complex models to explain the relationship between family life and work presented. Most likely, the problem of family life and work life is something obviously effects on family life. US National Research shows that 72 percent of men and 83 percent of women have an important conflict between expectations and their family roles and work experience (Leiter, MP, et al., 1994)

Overall, this study seeks to answer the question:

Is a positive relationship between job burnout Banking National Bank Tehran operators with the functions of their family there in 2015?

**Questions and research hypotheses:**

**Research questions:**

**Main questions:**

1. Is burn out among staff and family functions, there is a significant relationship?

**Secondary questions:**

1. Is burn out among staff and solving problems in family relationship there?
2. Is burn out among staff and communication skills of the family relationship there?
3. Is burn out among staff and their roles in the family relationship there?
4. Does the job burnout and emotional responses of family relationship there?
5. Does the employee's job burnout and emotional conflict in the family relationship there?
6. Is burn out among workers and social control of the family relationship there?

**Research hypotheses:**

**The main hypotheses:**

1. Burnout among staff and family functions there is a significant relationship.

**Secondary hypotheses:**

1. burnout among staff and solving problems in family relationship there.
2. Burnout among staff and communication skills in the family there is a significant relationship.
3. Burnout among staff and their roles in the family, there is a significant relationship.
4. Burnout among employees and their emotional responses to family ties there.
5. Burnout among employees and their emotional conflicts in the family, there is a significant relationship.
6. burnout among workers and social control of the family relationship there.

2. **Theory and literature:**
2.1 **Background research:**

No definition of family: the term refers to the relative unit (Pour Afkari 1997). In the nuclear, family consists of father, mother and children may be more extensive application of extended families that include both the grandfather, grandmother, aunt, uncle, aunt, etc. may also be adopted all of them as a social unit known act (Pour Afkari 1997)

In this study, the family consists of a group of people living together under one roof.

Family Functioning: McMaster model of family function, a behavioral model to consider six aspects of family functioning, problem solving, communication, roles, affective responsiveness, affective involvement and behavior control (also called a general after general function) each which of these dimensions on the continuum of efficiency and inefficiency.

Ernest Burgess today's most important family function to develop and maintain the character of its members Parsons knows emphasis on the social function of the stability and growth of children and adult personality. "Nayd Hart" In addition to the functions of reproduction, social and psychological protection functions to determine the social status of the family and reduces and eliminates tensions and also has reaffirmed housekeeping and leisure.

Sociologists in general terms and in the global framework of the basic tasks of the family are classified as follows:

Table (1): The basic functions of the family

| Davis | Gud | Marduk |
|---|---|---|
| Reproductive | Reproductive | Reproductive |
| conservation and protection | Physical conservation and protection | Economic |
| Socialization | Socialization | Education |
| Reproductive appropriate according to the conditions | Social control | Sexual |
| Placement and replacement | Placement and replacement | _____ |

Problem solving: the ability of families to solve problems and the steps that will follow to do this is when all the steps to be taken well, Entrepreneurs as possible has been done to solve the problem and solving the problem is that the proportion of families cannot even recognize it.

Communication: the ability of family members to exchange information (verbal information only)

Roles: The way the family in the distribution of tasks and so on.

Emotional response: the ability of family members to the appropriate emotional response, whether positive or negative emotions referred.

Affective involvement: the quality of love, attention and investment refers family members against each other.

Behavior control standards and describes the behavioral freedom.

In this study, the family functioning is what measures the Family Assessment Device.

Burnout: a syndrome consisting of emotional exhaustion, depersonalization and reduced progression among those people that can work directly with people seem (Maslach, 1982)

Approximately 20 years, the term burnout in psychology literature used and is now as important social problems and personal burnout known. (Farber, 1983, p. 17) And the relationship of people to their work and the problems because it leads to discontent and dissatisfaction as an important phenomenon in the years and is known to use the term burnout to the phenomenon of the 1970s in the US States became common, especially among people who were employed to work in human services. The first story Qerner (1981), an architect of job burnout and frustration about the mental problems that caused him to seek refuge in the forests of South Africa has been seen. (Bakker, A, et. Al, 2000)

Table (2): The main symptoms of burnout (based on moose studies, 1981)

| Job performance | Behavioral changes | Physical indicators |
|---|---|---|
| Reducing efficiency (spending a lot of time to do one thing at the same time, productivity is poor) | Increased irritability | Types of headaches |
| Loss of interest in work | Mood changes | Sleep disorder |
| Reduced capacity to maintain the effectiveness of performance in cases where a person is under stress. | Reduced capacity for failure tolerance | weight loss |
| Increased rigidity of thought (thought closed and inflexible) | Increased suspicion of others | Gastrointestinal disorders |
|  | More prepared to accept risk Trying to self-medication (drugs and narcotics) | Burnout |

Emotional exhaustion: emotional resources related to the reduction of the feeling that the mechanism is lacking resources.

Depersonalization: start a series of Dlsrdanh and cruel attitudes to their jobs.

Reduction of personal development: namely, that they cannot feel their work as well as it could do, do.

In this study, the Maslach Burnout is what measures the tool.

**2.2 Background Research:**

Studies and researches the issue that is close to this study was to summarize a few of them expressed as follows.

- In relation to the issue of burnout in our country for the first time in the Philippines (1992) study entitled "burnout and its relationship with coping styles of nurses employed in hospitals affiliated to Tehran" did. His results showed that a large number of nurses with job burnout grappling so that burnout was high at 75/28% of the cases related to the effects of demographic characteristics such as age, sex, marital status and shift and left the job burnout, There were significant. Results showed that nurses are the three dimensions of burnout in good condition, and it was found that young people and single vulnerable to burn out.
- Rafeie (1994) in a study entitled "Evaluation of burnout and its relationship with coping styles used by nurses in hospitals of Tehran Burn" got nurses who have used the method of high responsibility Burnout is high in nurses who have taken too much of problem-solving method Low levels of burnout and health care providers who have used methods to counter the negative emotion-oriented and high levels of burnout.
- Hasani (2002) in a study entitled "Evaluation of burnout math and art teachers in normal schools in Tehran," showed that none of his theories were not statistically significant and statistically based sample had to burn out the relationship between mathematics and art teacher burnout dimensions of emotional exhaustion, depersonalization and personal accomplishment, there was no significant difference. Burnout between married and single women teachers of mathematics and art the same three dimensions, there was no difference. The relationship between mathematics and art teacher burnout in three dimensions in terms of the relationship between field of study and teaching field, there was no significant difference, also in terms of experience, teaching area, away from the workplace.
- Saduk et al (1998) in a study entitled "The continuity of burnout in staff housing complexes for people with intellectual disabilities, 174 (127 women and 47 men) between 12 government agencies, and four non-governmental organizations examined they were the subjects completed a questionnaire Maslach burnout. The results showed that there was a significant difference between religious and nonreligious people. And religious subjects showed significantly lower burnout. Marital status was continuity with dimensions of burnout. The subjects who had been separated from his wife showed the highest level of burnout. People who had lost his wife also showed high levels of burnout. Among the scores of job burnout and working hours, job history, age, degree of mental disability client's significant difference was observed.
- Naagy et al (1992) in a paper entitled "A longitudinal study of burnout in teachers" Maslach Burnout Inventory over a five-year period in three innings on 100 teachers from elementary, junior high, high school was run. The study showed that high scores of emotional exhaustion and low score in the teachers' personal development gained. Most primary school teachers and high school teachers have to burn out. The results showed that not only the school, but the teacher should be the basis of an analysis unit at burnout is considered.

3. **Development of hypotheses and model:**

The above information is provided based on the concept that the design and sequencing of such a model should be considered causal order, and this inference and deduction based on research of the subject developed.

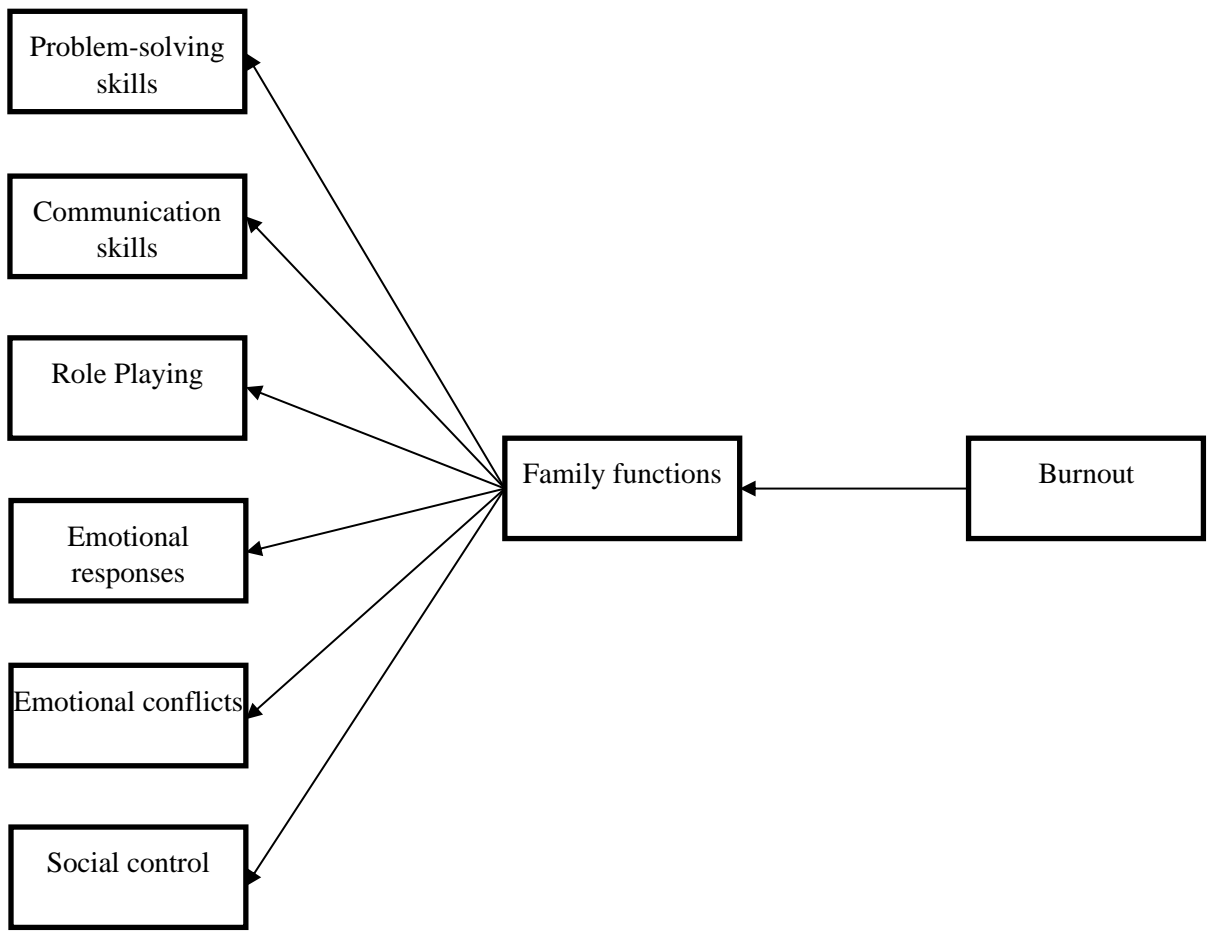

Figure (1): conceptual model

## 4. Methodology:

The method used descriptive statistical population used in this study consisted of 314 bank loan officers in branches of Melli Iran Bank of Tehran province Meanwhile, all the officials of the bank for more than five years of service in the Melli Iran Bank have branches in Tehran, and men are married and the average age of almost 38 years that make up the population of this study. The study sample consisted of 130 patients were selected based on Morgan. In this study, the questionnaires of the Maslach Burnout Inventory (MBI[1]) have 22 questions than final 90% alpha internal emotional exhaustion, depersonalization and low personal accomplishment Cronbach alpha of 79% (71%) and family assessment questionnaire (FAD[2]) has 60 questions that 70/0 internal finisher to solve the problem, the emotional response 51/0, 70/0 communication, emotional blend 69/0, 59/0 roles; behavior control is used 68/0.

Table (3): Correlations between measures of burnout MBI

| Lack of personal accomplishment | Depersonalization | Emotional exhaustion | |
|---|---|---|---|
| - | - | - | Emotional exhaustion |
| - | - | 0/77 | Depersonalization |
| - | -0/33 | -0/48 | Lack of personal accomplishment |

Table (4): Correlation between FAD scales

| Behavior control | Affective involvement | Emotional response | Roles | Relationship | Problem Solving | |
|---|---|---|---|---|---|---|
| - | - | - | - | - | - | Problem Solving |
| - | - | - | - | - | 0/688 | Relationship |
| - | - | - | - | 0/694 | 0/621 | Role's |

---

[1] Maslach Burnout Inventory
[2] Family Assessment Device

|   |   |   | 0/624 | 0/757 | 0/617 | Emotional response |
|---|---|---|---|---|---|---|
| - | - | - | 0/624 | 0/757 | 0/617 | Emotional response |
| - | - | 0/526 | 0/623 | 0/616 | 0/487 | Affective involvement |
| - | 0/446 | 0/442 | 0/593 | 0/536 | 0/563 | Behavior control |
| 0/629 | 0/591 | 0/694 | 0/688 | 0/782 | 0/743 | Overall performances |

## 5. Data analysis:

Table (5): Correlation between scales of family functioning and measures of burnout

| The level of significance | Lack of personal accomplishment | Depersonalization | Emotional exhaustion |   |
|---|---|---|---|---|
| 0/01 | +0/31 | -0/40 | -0/34 | Problem Solving |
| 0/01 | +0/30 | -0/40 | -0/34 | Relationship |
| 0/01 | 0/30 | -0/34 | -0/31 | Role's |
| 0/01 | 0/32 | -0/40 | -0/32 | Emotional response |
| 0/05 | 0/13 | -0/20 | -0/20 | Affective involvement |
| 0/05 | 0/30 | -0/20 | -0/30 | Behavior control |
| 0/01 | 0/32 | -0/40 | -0/40 | Overall performances |

Table (6): Correlation between scales of family functioning and burnout

| Significance level | Burnout | |
|---|---|---|
| 0/01 | -0/31 | Problem Solving |
| 0/01 | -0/35 | Relationship |
| 0/01 | -0/31 | Role's |
| 0/01 | -0/31 | Emotional response |
| 0/05 | -0/20 | Affective involvement |
| 0/05 | -0/22 | Behavior control |
| 0/01 | -0/38 | Overall performances |

"Burnout among staff and solving problems in family relationship there."

- The scale of the problem has a negative correlation with emotional exhaustion factor

$(r = -0/34)$ that level $P < 0/01$ the scale has a negative correlation with depersonalization factor $(r = -0/40)$ that level $P < 0/01$ and lack of positive correlation with personal accomplishment $(r = 0/31)$.

"Burnout among staff and communication skills of the family relationship there."

- The scale of the relationship has a negative correlation with emotional exhaustion factor

$(r = -0/34)$ that level $P < 0/01$ the scale has a negative correlation with depersonalization factor $(r = -0/40)$ that level $P < 0/01$ and lack of positive correlation with personal accomplishment $(r = 0/30)$.

"Burnout among staff and their role in the family relationship there."

- The scale of the roles has a negative correlation with emotional exhaustion factor $(r = -0/31)$ that level $P < 0/01$ the scale has a negative correlation with depersonalization factor $(r = -0/34)$ that level $P < 0/01$ and lack of positive correlation with personal accomplishment $(r = 0/30)$.

"Burnout among employees and emotional responses of family relationship there."

- The scale of the emotional responses has a negative correlation with emotional exhaustion factor $(r = -0/32)$ that level $P < 0/01$ the scale has a negative correlation with depersonalization factor $(r = -0/40)$ that level $P < 0/01$ and lack of positive correlation with personal accomplishment $(r = 0/30)$.

"Burnout among employees and their emotional conflicts in the family, there is a significant relationship."

- The scale of the Affective involvement has a negative correlation with emotional exhaustion factor $(r = -0/20)$ the scale has a negative correlation with depersonalization factor $(r = -0/20)$ that level $P < 0/05$ and lack of positive correlation with personal accomplishment $(r = 0/12)$.

"Burnout among workers and social control of the family relationship there."

- The scale of the Behavior control has a negative correlation with emotional exhaustion factor $(r = -0/30)$ that level $P < 0/05$ the scale has a negative correlation with depersonalization factor $(r = -0/20)$ that level $P < 0/01$ and lack of positive correlation with personal accomplishment $(r = 0/30)$.
- The scale of the Overall performances has a negative correlation with emotional exhaustion factor $(r = -0/40)$ the scale has a negative correlation with depersonalization factor $(r = -0/40)$ that level $P < 0/01$ and lack of positive correlation with personal accomplishment $(r = 0/32)$.

6. **Conclusion:**

The results indicate that the hypothesis that included the relationship between burnout and 6, the family functioning, problem solving, communication, roles, affective responsiveness, affective fusion, there was a significant relationship between behavior and the correlation was negative. The burnout is high; the functions of the family will be in trouble. Of course, the results of this study with results in similar studies on job burnout and family functions confirmed. This means that the research and Burke & Greenglass (2001) It conducted a burnout feature is contagious and can have negative impacts on people's lives also leave behind. Bolger and others (1989) note that stress is contagious quality and other aspects of life spreads and Cooper (1992), the idea that family issues are clearly working life and working life and family life affects confirmed.

**Offers:**

1. Encourage employees to give high importance to your health with the aim of preventing burnout and coping and treatment through:
- Access to MBI by filling it to your bank site with a guide to appraisal and preparation and delivery of services to people at various levels, including clinical services, counseling services, peer education and virtual education;
- Taking part of the site and one page of every issue of the bank to provide information on methods of stress reduction for one year continuously;
2. Encouraging employees to exercise by facilitating access to sports services aimed at reducing burnout and prevention of it by:
- Assign a percentage of the cost of health findings on physical activity;
- Training exercise to reduce physical fatigue and reduce the negative effects of computer use Brochures, Leaflets, educational books, movies and ...
- The payment of overtime for participating in exercise classes;
3. According to the results of comparative tests mean demographic characteristics of individuals and their role in job burnout following suggestions are offered:
- Assistance to single employees to provide marriage and family circumstances, by providing facilities and loans, in-kind assistance.